\documentclass{nature_mod}

\usepackage{url}
\usepackage{epsfig}

\newcommand{\apj}{Astrophys. J.}
\newcommand{\apjl}{Astrophys. J.}
\newcommand{\apjs}{Astrophys. J.}
\newcommand{\aap}{Astron. Astrophys.}
\newcommand{\mnras}{Mon. Not. R. Astron. Soc.}
\newcommand{\nat}{Nature}
\newcommand{\araa}{Ann. Rev. Astron. Astrophys.}

\usepackage{amssymb,amsmath,caption}

\newcommand{\frb}{FRB~121102}

\long\def\symbolfootnote[#1]#2{\begingroup%
\def\thefootnote{\fnsymbol{footnote}}\footnote[#1]{#2}\endgroup} 

\title{A Repeating Fast Radio Burst}

\author{
L.~G.~Spitler$^{1}$, 
P.~Scholz$^{2}$,
J.~W.~T.~Hessels$^{3,4}$, 
S.~Bogdanov$^{5}$, 
A.~Brazier$^{6,7}$, 
F.~Camilo$^{5,8}$, 
S.~Chatterjee$^{6}$,
J.~M.~Cordes$^{6}$, 
F.~Crawford$^{9}$, 
J.~Deneva$^{10}$,
R.~D.~Ferdman$^{2}$, 
P.~C.~C.~Freire$^{1}$, 
V.~M.~Kaspi$^{2}$, 
P.~Lazarus$^{1}$, 
R.~Lynch$^{11,12}$, 
E.~C.~Madsen$^{2}$,
M.~A.~McLaughlin$^{12}$, 
C.~Patel$^{2}$, 
S.~M.~Ransom$^{13}$, 
A.~Seymour$^{14}$, 
I.~H.~Stairs$^{15,2}$, 
B.~W.~Stappers$^{16}$, 
J.~van~Leeuwen$^{3,4}$ \&
W.~W.~Zhu$^{1}$
}

\begin{document}

\maketitle

\begin{affiliations}
 \item Max-Planck-Institut f\"ur Radioastronomie, Auf dem H\"{u}gel 69, B-53121 Bonn, Germany
 \item Department of Physics and McGill Space Institute, McGill University, 3600 University St., Montreal, QC H3A 2T8, Canada
 \item ASTRON, Netherlands Institute for Radio Astronomy, Postbus 2, 7990 AA, Dwingeloo, The Netherlands
 \item Anton Pannekoek Institute for Astronomy, University of
   Amsterdam, Science Park 904, 1098 XH Amsterdam, The Netherlands
 \item Columbia Astrophysics Laboratory, Columbia University,  New York, NY 10027, USA
 \item Department of Astronomy and Space Sciences, Cornell University, Ithaca, NY 14853, USA
 \item Cornell Center for Advanced Computing, Cornell University, Ithaca, NY 14853, USA
 \item Square Kilometre Array South Africa, Pinelands, 7405, South Africa
 \item Department of Physics and Astronomy, Franklin and Marshall College, Lancaster, PA 17604-3003, USA
 \item Naval Research Laboratory, 4555 Overlook Ave SW, Washington, DC 20375, USA
 \item National Radio Astronomy Observatory, PO Box 2, Green Bank, WV, 24944, USA
 \item Department of Physics and Astronomy, West Virginia University, Morgantown, WV 26506, USA
 \item National Radio Astronomy Observatory, Charlottesville, VA 22903, USA
 \item Arecibo Observatory, HC3 Box 53995, Arecibo, PR 00612, USA
 \item Department of Physics and Astronomy, University of British Columbia, 6224 Agricultural Road Vancouver, BC V6T 1Z1, Canada
 \item Jodrell Bank Centre for Astrophysics, School of Physics and Astronomy, University of Manchester, Manchester, M13 9PL, UK
\end{affiliations}

\bigskip

\begin{abstract}
  Fast Radio Bursts are millisecond-duration astronomical radio pulses
  of unknown physical origin that appear to come from extragalactic
  distances\cite{lbm+07,tsb+13,bb14,sch+14,pbb+15,rsj15,cpk+15,mls+15a}.
  Previous follow-up observations have failed to find additional
  bursts at the same dispersion measures (i.e. integrated column
  density of free electrons between source and telescope) and sky
  position as the original detections\cite{pjk+15}.  The apparent
  non-repeating nature of the fast radio bursts has led several
  authors to hypothesise that they originate in cataclysmic
  astrophysical events\cite{fr14}.  Here we report the detection of
  ten additional bursts from the direction of \frb, using the 305-m
  Arecibo telescope.  These new bursts have dispersion measures and
  sky positions consistent with the original burst\cite{sch+14}.  This
  unambiguously identifies \frb\ as repeating and demonstrates that
  its source survives the energetic events that cause the bursts.
  Additionally, the bursts from \frb\ show a wide range of spectral
  shapes that appear to be predominantly intrinsic to the source and
  which vary on timescales of minutes or shorter.  While there may be
  multiple physical origins for the population of fast radio bursts,
  the repeat bursts with high dispersion measure and variable spectra
  specifically seen from \frb\ support models that propose an origin
  in a young, highly magnetised, extragalactic neutron
  star\cite{cw15,pc15}.
\end{abstract}

\clearpage


\frb\ was discovered\cite{sch+14} in the PALFA survey, a deep search
of the Galactic plane at 1.4\,GHz for radio pulsars and fast radio
bursts (FRBs) using the 305-m William E. Gordon Telescope at the
Arecibo Observatory and the 7-beam Arecibo L-band Feed Array
(ALFA)\cite{cfl+06,lbh+15}.  The observed dispersion measure (DM) of
the burst is roughly three times the maximum value expected along this
line of sight in the NE2001 model\cite{cl02} of Galactic electron
density, i.e. $\beta_{\rm DM} \equiv \rm{DM_{FRB}/DM^{Gal}_{Max}} \sim
3$, suggesting an extragalactic origin.

\begin{figure}
\centerline{\includegraphics[width=0.9\textwidth]{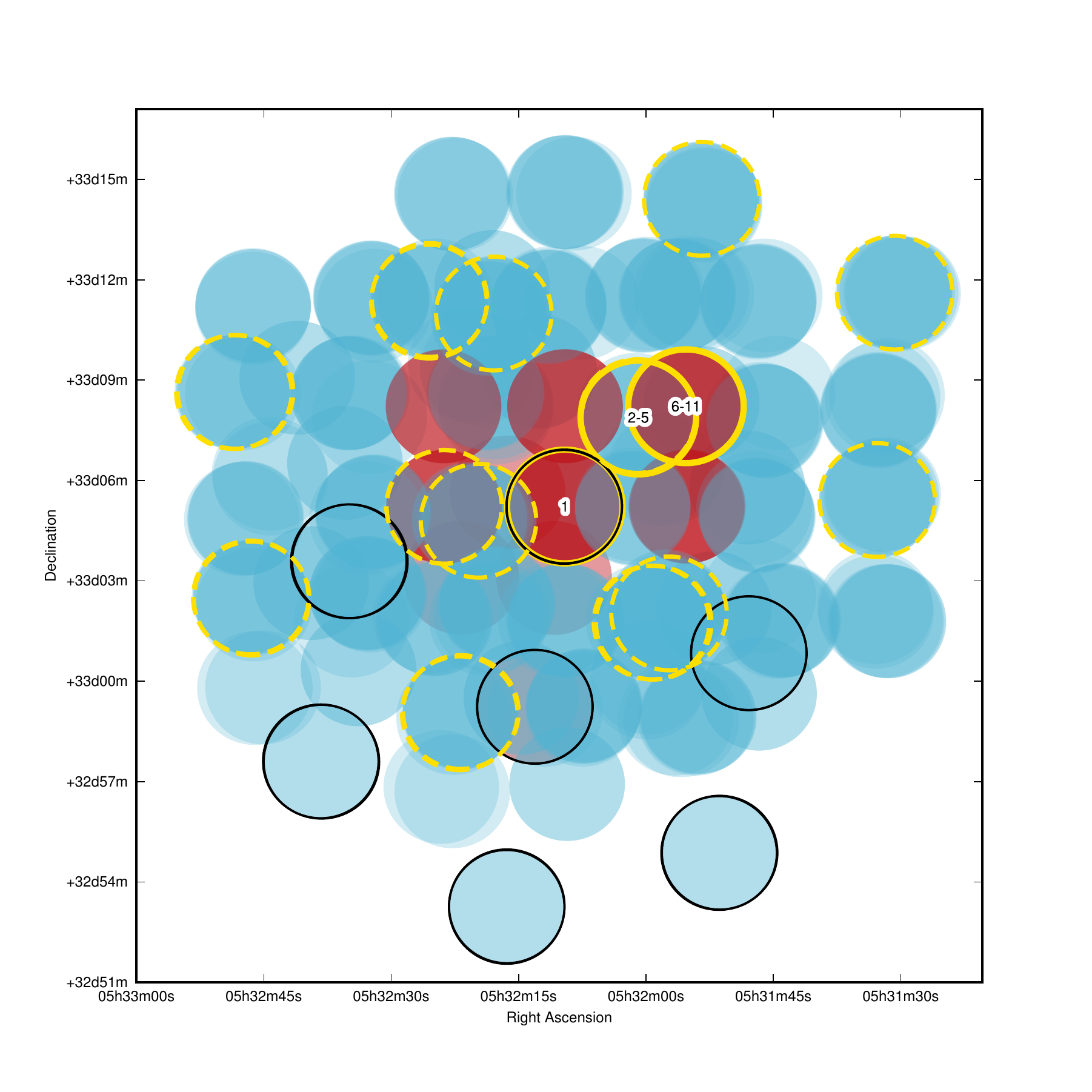}}
\caption{{\bf Discovery and follow-up detections of \frb.}  For each
  seven-beam ALFA pointing, the central and outer six beams are shown
  in red and blue, respectively (see Extended Data
  Table~\ref{tab:observations} and Table~\ref{tab:gridpoints}).  The
  circles indicate the $\sim 3.5^{\prime}$ half-power widths of the
  beams at 1.4\,GHz.  Darker shading indicates sky positions with
  multiple grid observations at roughly the same position.  The
  initial discovery pointing\cite{sch+14} and second survey
  observation are outlined in black (these overlap).  Beam positions
  in which bursts were detected are outlined in thick, solid gold
  (dashed for the other six beams from the same pointing) and the
  corresponding burst identifier numbers (Table~\ref{tab:bursts}) are
  also given.
\label{fig:gridding}}
\end{figure}

Initial Arecibo follow-up observations were limited in both dwell time
and sky coverage and resulted in no detection of additional
bursts\cite{sch+14}.  In 2015 May and June we carried out more
extensive follow-up using Arecibo, covering a $\sim 9^{\prime}$ radius
with a grid of six ALFA pointings around the then-best sky position of
\frb\ (Figure~\ref{fig:gridding} and Extended Data
Table~\ref{tab:observations} and \ref{tab:gridpoints}).  As described
in the Methods, high-time-resolution, total intensity spectra were
recorded, and the data were processed using standard radio-frequency
interference (RFI) excision, dispersion removal, and
single-pulse-search algorithms implemented in the PRESTO software
suite and associated data reduction
pipelines\cite{ran01,kkl+15,lbh+15}.

\begin{figure}
\centerline{\includegraphics[scale=0.65]{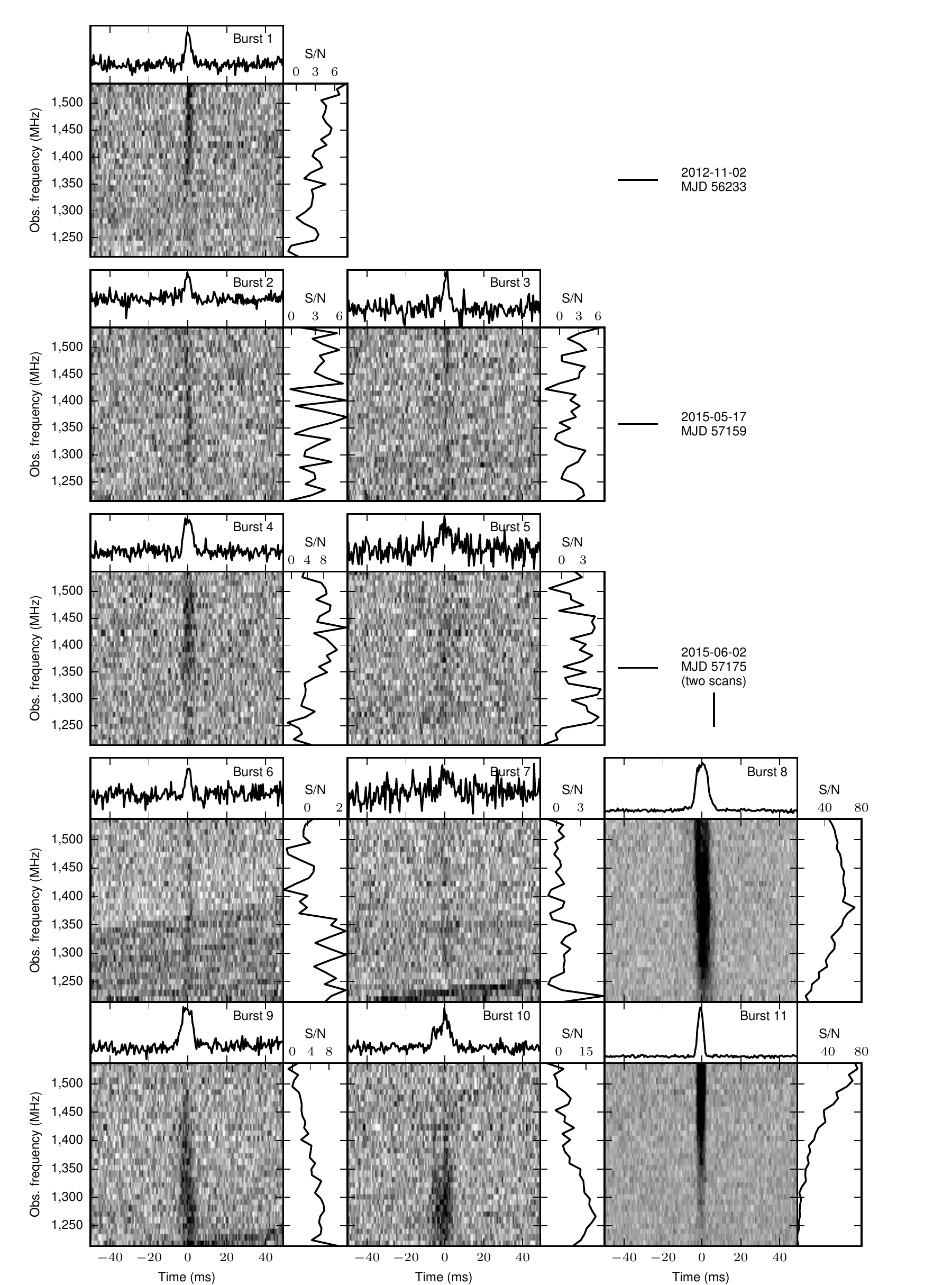}}
\caption{{\bf \frb\ burst morphologies and spectra.}  The central
  greyscale (linearly scaled) panels show the total intensity versus
  observing frequency and time, after correcting for dispersion to
  DM=559\,pc\,cm$^{-3}$. The data are shown with 10\,MHz and 0.524\,ms
  frequency and time resolution, respectively. The diagonal striping
  at low radio frequencies for Bursts 6, 7 and 9 is due to RFI that is
  unrelated to \frb.  The upper sub-panels are burst profiles summed
  over all frequencies. The band-corrected burst spectra are shown in
  the right sub-panels.  The signal-to-noise scales for the spectra
  are shown on each sub-panel. All panels are arbitrarily and
  independently scaled.
\label{fig:bursts}}
\end{figure}

We detected 10 additional bursts from \frb\ in these observations.
The burst properties, and those of the initial \frb\ burst, are listed
in Table~\ref{tab:bursts}.  The burst intensities are shown in
Figure~\ref{fig:bursts}.  No other periodic or single-pulse signals of
a plausible astrophysical origin were detected at any other DM.  Until
the source's physical nature is clear, we continue to refer to it as
\frb\ and label each burst chronologically starting with the original
detection.

The ten newly detected bursts were observed exclusively in two
adjacent sky positions of the telescope pointing grid located $\sim
1.3^{\prime}$ apart (Figure~\ref{fig:gridding} and Extended Data
Table~\ref{tab:observations}).  The unweighted average J2000 position
from the centres of these two beams is right ascension $\alpha =
05^h31^m58^s$, declination $\delta = +33^d08^m04^s$, with an
uncertainty radius of $\sim 3^{\prime}$.  The corresponding Galactic
longitude and latitude are $ l = 174^{\circ}\!\! .89, b =
-0.\!\!^{\circ}23$.  This more accurate position is $3.7^{\prime}$
from the beam centre of the discovery burst\cite{sch+14}, meaning that
\frb\ Burst 1 was detected well off axis, as originally concluded.

The measured DMs of all 11 bursts are consistent to within the
uncertainties, and the dispersion indices (dispersive delay $\Delta t
\propto \nu^{-\xi}$) match the $\xi = 2.0$ value expected for radio
waves traveling through a cold, ionised medium.  This is strong
evidence that a single astronomical source is responsible for the
events.  In addition, the $\sim 0.002$ DM index uncertainty we
calculate for Burst 11 (see Methods) is slightly less than that
reported for FRB~110523\cite{mls+15a}, making this the most precise
determination of dispersion index for any FRB thus far. The upper
bound on the dispersion index is identical to that of
FRB~110523\cite{mls+15a}, and hence, following the same arguments used
there, Burst 11 provides a similar lower limit of 10 AU for the size
of the dispersive region.

The 11 bursts have peak flux densities $gS_{1400} \sim 0.02 - 0.3$\,Jy
at 1.4\,GHz, where $g$ is the antenna gain at the source's unknown
location in the beam normalised to unit amplitude on the beam axis.
The other known FRBs typically have order-of-magnitude higher peak
flux densities of $gS_{1400} \sim 0.2 - 2$\,Jy.  The wide range of
flux densities seen at Arecibo, some near the detection threshold,
suggests that weaker bursts are also produced, likely at a higher
rate.  The rate of burst detections is $\sim$3 hr$^{-1}$ for bursts
with $gS_{1400} \gtrsim 20$\,mJy over all observations in which an
ALFA beam was within $3.5^\prime$ of the improved position.  We note,
however, that the bursts appear to cluster in time with some observing
sessions showing multiple bright bursts and others showing none.

The observed burst full widths at half maximum are $w_{50} = 2.8 -
8.7$\,ms, which are consistent with the $w_{50} = 1.3 - 23.4$\,ms
widths seen from other FRBs.  No clear evidence for scatter broadening
was seen in any of the bursts. Bursts 8 and 10 show double-peaked
profiles, which has also been seen in FRB~121002\cite{cpk+15}.
Furthermore, the morphologies of Bursts 8 and 10 evolve smoothly with
frequency.

Within our observing band ($1.214 -Ð 1.537$ GHz) the burst spectra are
remarkably variable.  Some are brighter toward higher frequencies, as
in the initial discovery, Burst 1, while others are brighter toward
lower frequencies.  The spectra of Bursts 8 and 10 are not monotonic.
The detections of Bursts 6--11 exclusively in Beam 0 of the ALFA
receiver (see Extended Data Table 1) means that the bursts must have
been detected in the main beam and not in a side-lobe.  While the
frequency-dependent shape of the main beam attenuates the bursts'
intrinsic spectra at higher frequencies if the source is
off-axis\cite{sch+14}, this bias is either not large enough or in the
wrong direction to cause the observed spectral variability of Bursts
6--11.  Given our improved position, Burst 1 is consistent with its
detection in a side-lobe, which, unlike in the main beam, could have
caused attenuation of the spectrum at lower frequencies.  This
spectral volatility is reflected by the wide range of spectral indices
$\alpha \sim -10$ to +14 obtained from fitting a power-law model
($S_\nu \propto \nu^{\alpha}$, where $S_\nu$ is the flux density at
frequency $\nu$) to burst spectra (Table~\ref{tab:bursts}).

There is no evidence for fine-scale diffractive interstellar
scintillation, most likely because it is unresolved by our limited
spectral resolution.  In principle, the spectra could be strongly
modulated if the source is multiply imaged by refraction in the
interstellar medium\cite{wc87} or by gravitational lensing.  However,
the splitting angle between sub-images required to produce spectral
structure across our band ($\ll 1$ milli-arcsecond) is much smaller
than the expected diffraction angle from interstellar plasma
scattering.  The fine-scaled diffraction structure in the spectrum
will therefore wash out the oscillation.  Lastly, positive spectral
indices could also be explained by free-free absorption at the
source\cite{kon15}, but this is ruled out by the large spectral
differences among bursts.  We therefore conclude that the spectral
shapes and variations are likely to be predominantly intrinsic to the
source.

An analysis of the arrival times of the bursts did not reveal any
statistically significant periodicity (see Methods).  If the source
has a long period ($\gtrsim 1$\,s), then it is likely emitting at a
wide range of rotational phases, which is not uncommon for
magnetars\cite{crh+06}, making a convincing period determination
difficult.  Due to the small number of detected bursts, we are not
sensitive to periodicities much shorter than $\sim 100$\,ms.

Repeat bursts rule out models involving cataclysmic events -- such as
merging neutron stars\cite{hl01} or collapsing supra-massive neutron
stars\cite{fr14}.  Bursts from Galactic flare stars have been proposed
as a model for FRBs with the DM excess originating in the stellar
corona\cite{lsm14}.  However, temporal density variations in the
corona should produce bursts with varying DMs, which we do not
observe.  Planets orbiting in a magnetised pulsar wind may produce a
millisecond-duration burst once per orbital period\cite{mz14};
however, the observed intra-session separations of our bursts (23 to
572\,s) are too short to correspond to orbital periods.  Repeated
powerful radiative bursts are associated with magnetars, and indeed
giant flares from the latter have been suggested as a FRB
source\cite{lyu14,pc15,kon15}.  However, no Galactic magnetar has been
seen to emit more than a single giant flare in over four decades of
monitoring, arguing against a magnetar giant flare origin for \frb.
Magnetars have been observed to exhibit repeating bright radio
pulses\cite{crh+06}, however not yet at the energy scale implied if
\frb\ is more than several hundred kpc away.

Giant pulse emission from an extragalactic pulsar remains a plausible
model\cite{cw15}.  The most prominent giant pulses are from the Crab
pulsar, which has a large spin-down energy loss rate.  Spectral
indices calculated from wideband measurements of giant pulses from the
Crab pulsar\cite{kss10} have a broad distribution ranging from $\alpha
\sim -15$ to +10, as well as frequency ``fringes'' -- i.e. a banded
structure to the emission brightness as a function of
frequency\cite{he07}.  These fringes have characteristic widths of a
few hundred MHz, and we speculate that -- given our 322-MHz observing
bandwidth -- a similar phenomenon could create the spectral
variability we have seen in \frb.  The double-peaked nature of some
\frb\ bursts is also possible in the giant pulse model\cite{cpk+15},
and the evolution of these burst morphologies with frequency could
imply rapid spectral variation between consecutive (sub-)pulses only
milliseconds apart.

The low Galactic latitude and relatively small $\beta_{\rm DM}$ of
\frb\ compared with other FRBs raises the question whether it is
genuinely extragalactic in origin (see also Methods).  However, no
H$\alpha$ or HII regions are seen in archival data along the line of
sight to \frb, as might be expected for an intervening ionised
nebula\cite{sch+14} that can give $\beta_{\rm DM} \gg 1$.
Furthermore, a detailed multi-wavelength investigation, which searched
for a compact nebula in a sky region that includes the refined
position presented here, concluded that \frb's high DM cannot be
explained by unmodelled Galactic structure along the line of sight and
that \frb\ must therefore be extragalactic\cite{kon15}.  Conclusively
establishing that \frb\ is extragalactic will require arcsecond
localisation and association with a host galaxy.  The repeating nature
of the bursts facilitates such localisation with a radio
interferometer.

While the \frb\ bursts share many similarities to the FRBs detected
using the Parkes\cite{lbm+07,tsb+13,bb14,pbb+15,rsj15,cpk+15} and
Green Bank\cite{mls+15a} telescopes, it is unclear whether \frb\ is
representative of all FRBs.  The 10 bursts from \frb\ in 2015 were
detected near the best-known position in 3\,hrs of observations.  In
contrast, follow-up observations of the Parkes FRBs, again using the
Parkes telescope, range in total time per direction from a few
hours\cite{pjk+15} to almost 100 hours\cite{lbm+07} and have found no
additional bursts.  Arecibo's $> 10\times$ higher sensitivity may
allow detection of a broader range of the burst-energy distribution of
FRBs, thus increasing the chances of detecting repeated bursts; for
example, of the 11 bursts from \frb, Parkes may have been capable of
detecting only Bursts 8 or 11.  More sensitive observations of the
Parkes FRBs may therefore show that they also sporadically repeat.

Alternatively, \frb\ may be fundamentally different from the FRBs
detected at Parkes and Green Bank.  As was the case for supernovae and
gamma-ray bursts, multiple astrophysical processes may be required to
explain a diversity of observational properties of FRBs.\\ 
\\




\begin{addendum}

\item We thank the staff of the Arecibo Observatory, and in particular
  A.~Venkataraman, H.~Hernandez, P.~Perillat and J.~Schmelz, for their
  continued support and dedication to enabling observations like those
  presented here. We also thank our commensal observing partners from
  the Arecibo ``Zone of Avoidance'' team, in particular T.~McIntyre
  and T.~Henning.  We also thank M.~Kramer for helpful suggestions.
  The Arecibo Observatory is operated by SRI International under a
  cooperative agreement with the National Science Foundation
  (AST-1100968), and in alliance with Ana G.~M\'{e}ndez-Universidad
  Metropolitana, and the Universities Space Research Association.
  These data were processed using the McGill University High
  Performance Computing Centre operated by Compute Canada and Calcul
  Qu\'{e}bec.  The National Radio Astronomy Observatory is a facility
  of the National Science Foundation operated under cooperative
  agreement by Associated Universities, Inc.  The research leading to
  these results has received funding from the European Research
  Council (ERC) under the European Union's Seventh Framework Programme
  (FP7/2007-2013).  L.G.S., P.C.C.F. and P.L. gratefully acknowledge
  financial support from the ERC Starting Grant BEACON under contract
  no. 279702.  J.W.T.H. is an NWO Vidi Fellow and also gratefully
  acknowledges funding for this work from ERC Starting Grant DRAGNET
  under contract no. 337062.  Work at Cornell (J.M.C., S.C., A.B.) was
  supported by NSF grants AST-1104617 and AST-1008213.  V.M.K. holds
  the Lorne Trottier Chair in Astrophysics and Cosmology and a
  Canadian Research Chair in Observational Astrophysics and received
  additional support from NSERC via a Discovery Grant and Accelerator
  Supplement, by FQRNT via the Centre de Recherche Astrophysique de
  Qu\'ebec, and by the Canadian Institute for Advanced Research.
  J.v.L. acknowledges funding for this research from an ERC
  Consolidator Grant under contract no. 617199.  Pulsar research at
  UBC is supported by an NSERC Discovery Grant and by the Canadian
  Institute for Advanced Research.

\item[Author Contributions] L.G.S. and J.W.T.H. led the design and
  execution of the observing campaign described here.  P.S. performed
  the analysis that discovered the radio bursts.  More detailed
  analysis of the signal properties was done by L.G.S., P.S., S.M.R.,
  M.A.M., J.W.T.H., S.C. and J.M.C..  L.G.S., J.W.T.H., P.S. and
  V.M.K. led the writing of the manuscript.  All authors contributed
  significantly to the interpretation of the analysis results and to
  the final version of the manuscript.

\item[Competing Interests] The authors declare that they have no
  competing financial interests.

\item[Correspondence] Correspondence and requests for materials should
  be addressed to J.W.T.H.~(email: J.W.T.Hessels@uva.nl).

\end{addendum}

\clearpage


\begin{table}
    \begin{center}
    \scriptsize
    \caption{\footnotesize {\bf Properties of detected bursts.}
      Uncertainties are the 68\% confidence interval, unless otherwise stated.
    \label{tab:bursts} }
    \begin{tabular}{lllllllll}
    \hline
    \hline
    No. & Barycentric         & Peak Flux        & Fluence      & Gaussian  & Spectral & DM               \\
        & Peak Time (MJD)$^a$ & Density (Jy)$^b$ & (Jy\,ms)$^b$ & Width$^c$ (ms) & Index$^d$    & (pc\,cm$^{-3}$)$^e$ \\
    \hline
    1 & 56233.282837008 & 0.04 & 0.1  & 3.3$\pm$0.3  &  8.8$\pm$1.9  & 553$\pm$5$\pm$2  \\
    2 & 57159.737600835 & 0.03 & 0.1  & 3.8$\pm$0.4  &  2.5$\pm$1.7  & 560$\pm$2$\pm$2   \\
    3 & 57159.744223619 & 0.03 & 0.1  & 3.3$\pm$0.4  &  0.9$\pm$2.0  & 566$\pm$5$\pm$2  \\
    4 & 57175.693143232 & 0.04 & 0.2  & 4.6$\pm$0.3  &  5.8$\pm$1.4  & 555$\pm$1$\pm$2  \\
    5 & 57175.699727826 & 0.02 & 0.09 & 8.7$\pm$1.5  &  1.6$\pm$2.5 & 558$\pm$6$\pm$4  \\
    6 & 57175.742576706 & 0.02 & 0.06 & 2.8$\pm$0.4  &                          & 559$\pm$9$\pm$1  \\
    7 & 57175.742839344 & 0.02 & 0.06 & 6.1$\pm$1.4  & -3.7$\pm$1.8 & & \\
    8 & 57175.743510388 & 0.14 & 0.9  & 6.6$\pm$0.1  &                            & 556.5$\pm$0.7$\pm$3  \\
    9 & 57175.745665832 & 0.05 & 0.3  & 6.0$\pm$0.3  &  -10.4$\pm$1.1 & 557.4$\pm$0.7$\pm$3  \\
    10& 57175.747624851 & 0.05 & 0.2  & 8.0$\pm$0.5  &                            & 558.7$\pm$0.9$\pm$4 \\
    11& 57175.748287265 & 0.31 & 1.0  & 3.06$\pm$0.04 &  13.6$\pm$0.4 & 556.5$\pm$0.1$\pm$1 \\
    \hline
    \hline
    \newline
    \end{tabular}
    \\$^a$ Arrival time corrected to the solar-system barycentre and referenced to infinite frequency (i.e the time delay due to dispersion is removed). \\
    $^b$ Lower limit since it assumes burst is detected at the centre of the beam (i.e.\ an assumed gain of 10 K/Jy yielding a system equivalent flux density of 3 Jy).\\
    $^c$ Widths are the full-width at half maximum. \\
    $^d$ Bursts 8 and 10 are not well-fit by a power-law model. Burst 6 is too corrupted by RFI to include.\\
    $^e$ Quoted errors are, in order, statistical and systematic (see Methods). Burst 7 was too weak and corrupted by RFI to include. \\
    \end{center}
\end{table}


\clearpage

\clearpage

\clearpage


\section*{Methods}

\subsubsection*{Positional gridding strategy and burst localisation}

Extended Data Table~\ref{tab:observations} summarises all the
observations made in both the initial 2013 follow-up and 2015 May/June
observations (project code p2886).  The ALFA Beam 0 pointing positions
in J2000 equatorial coordinates are summarised in Extended Data
Table~\ref{tab:gridpoints}.  In p2030/p2886 observations, the major
axis of the ALFA receiver was rotated $19^{\circ}$/$90^{\circ}$ with
respect to North\cite{sch+14}.

In 2015 May/June, we searched for additional bursts from \frb\ using a
grid of six pointings using the seven-beam ALFA receiver to cover a
generous $\sim 9^{\prime}$ radius around the discovery beam position
and side-lobe.  The ALFA receiver was aligned East-West to optimise the
sky coverage for this specific purpose.  The centre beams of the six
grid pointings are shown in red in Figure~\ref{fig:gridding}, and the
six outer ALFA beams are shown in blue.  Each grid pointing position
was observed at least four times for $\sim 1000$\,s.  The beam
positions of the discovery observation and 2013 follow-up
gridding\cite{sch+14} with ALFA (in that case rotated $19^{\circ}$
with respect to North) are also indicated using the same colour
scheme.  The outer six ALFA beams in the multiple grid observations
are only at roughly the same position because the projection of the
ALFA beams on the sky depends on the position of the telescope feed
with respect to the primary reflecting dish, and these do not overlap
perfectly between independent observations.

Two bursts on May 17 (Bursts 2 and 3) and two on June 2 (Bursts 4 and
5) were detected at a single grid position: FRBGRID2b in ALFA Beam 6,
which had positions of $\alpha = 05^h32^m01^s$, $\delta =
+33^d07^m56^s$ and $\alpha = 05^h32^m01^s$, $\delta = +33^d07^m53^s$
(J2000) at the two epochs -- i.e. only a few arcseconds apart.  Six
more bursts (Bursts 6--11) were detected on June 2 at a neighbouring
grid position, FRBGRID6b in ALFA Beam 0, $\sim 1.3^{\prime}$ away at
$\alpha = 05^h31^m55^s$, $\delta = +33^d08^m13^s$.  In all cases
bursts were detected in only one beam of the seven-beam ALFA receiver
at any given time.  This shows that the bursts must originate beyond
Arecibo's Fresnel length of $\sim 100$\,km\cite{kon15}.

The intermittency of \frb\ makes accurate localisation more
challenging.  Nonetheless, the detection in adjacent grid positions is
informative, and to refine the position of \frb, we simply take the
average position between FRBGRID2b ALFA Beam 6 and FRBGRID6b ALFA Beam
0, which gives: $\alpha = 05^h31^m58^s$, declination $\delta =
+33^d08^m04^s$ (J2000) and, equivalently, Galactic longitude and
latitude $ l = 174^{\circ}\!\!  .89, b = -0.\!\!^{\circ}23$.  The
approximate uncertainty radius of $\sim 3^{\prime}$ is based on the
amount of overlap between the two detection beam positions and the
ALFA beam width at half power, which is $\sim 3.5^{\prime}$.  The
distance from the initially reported Burst 1 position is
$3.7^{\prime}$, consistent with the interpretation that this burst was
detected in a side-lobe.  Though \frb\ bursts have been detected in
beams with different central sky positions, all detections are
consistent with a well defined sky position when one considers the
imprint of the ALFA gain pattern on the sky during each
observation\cite{sch+14}.

\subsubsection*{Galactic versus extragalactic interpretation}

Noteworthy is the fact that \frb\ lies directly in the Galactic plane,
whereas the other claimed FRBs lie predominantly at high Galactic
latitudes.  The PALFA survey is only searching in the Galactic plane,
however, and no comparable FRB survey at 1.4\,GHz with Arecibo has
been done at high Galactic latitudes.  Therefore, this difference may
simply be a consequence of where Arecibo has most deeply searched for
FRBs and does not necessarily suggest that \frb\ is of Galactic
origin.  Furthermore, \frb\ was found in the Galactic anti-centre
region of the PALFA survey, whereas searches in the inner-Galaxy
region have thus far found no FRBs\cite{lbh+15}.  This may be because
the Galactic foregrounds in the anti-centre region are comparatively
low, so the deleterious effects of DM smearing and scattering, which
may reduce our sensitivity to FRBs, are less important in the outer
Galaxy than inner Galaxy.

The low Galactic latitude of \frb\ also contributes to its low DM
excess factor $\beta_{\rm DM} \sim 3$ compared to the $\beta_{\rm DM}
\sim 1.2 - 40$ range seen for the other 15 FRBs in the literature.
Only FRB~010621\cite{kskl12}, with $\beta_{\rm DM} \sim 1.2$, has a
lower $\beta_{\rm DM}$ than \frb, and it has been proposed to be
Galactic\cite{bm14}.  We note, however, that six of 16 FRBs have DMs
comparable to or lower than \frb.  Furthermore, its total Galactic DM
excess $\rm{DM_{FRB} - DM^{Gal}_{Max}} \sim 370$\,pc\,cm$^{-3}$ is
larger than that of the first-discovered FRB\cite{lbm+07}.  Lastly,
within a generous 20-degree radius of \frb, the highest-DM pulsar
known is the millisecond pulsar PSR~J0557+1550\cite{skl+15} (also a
PALFA discovery), which has DM$ = 103$\,pc\,cm$^{-3}$ and $\beta_{\rm
  DM} = 0.6$, as well as the highest DM-inferred distance\cite{cl02}
of any pulsar in this region, $d = 5.7$\,kpc.  \frb's DM is clearly
anomalous, even when compared to this distant Galactic anti-centre
pulsar.  At an angular offset of 38 degrees, we note the existence of
PSR~J0248+6021, with DM$ = 370$\,pc\,cm$^{-3}$ and $\beta_{\rm DM} =
1.8$.  While the DM of this young, 217-ms pulsar is in excess of the
maximum Galactic contribution in the NE2001 model\cite{cl02}, this can
be explained by its location within the dense, giant HII region W5 in
the Perseus arm\cite{tpc+11} at a distance of 2\,kpc.  A similar
association for \frb\ has been sought to explain its $\beta_{\rm DM}
\sim 3$, but multi-wavelength investigations have as yet found no
unmodelled Galactic structure\cite{sch+14,kon15}.  In summary, \frb's
comparatively low $\beta_{\rm DM}$ does not strongly distinguish it
from other FRBs, or necessarily suggest it is more likely to be
Galactic.

\subsubsection*{Observations and search processing}

Here we provide a brief description of the Arecibo Mock spectrometer
data and search pipeline\cite{lbh+15} used for our follow-up
observations of \frb.  The 1.4-GHz data were recorded with the Mock
spectrometers, which cover the full ALFA receiver bandwidth in two
subbands.  Each 172-MHz subband was sampled with 16 bits, a time
resolution of 65.5\,$\mu$s, and frequency resolution of 0.34\,MHz in
512 channels.  The data were later converted to 4-bit samples to
reduce the data storage requirements.  Before processing, the two
subbands were combined into a single band of 322\,MHz (accounting for
frequency overlap between the two subbands), which was centred at
1375\,MHz and spans (1214.3 -- 1536.7 MHz).

We used the PALFA PRESTO-based\cite{ran01} search
pipeline\cite{lbh+15} to search for astrophysical signals in the
frequency and time domains.  These data were processed using the
McGill University High Performance Computing Centre operated by
Compute Canada and Calcul Qu\'{e}bec.  The presence of radio frequency
interference (RFI) can have a detrimental effect on our ability to
detect bursts.  We therefore applied PRESTO's {\tt rfifind} to
identify contaminated frequency channels and time blocks.  Flagged
channels and time blocks were masked in subsequent analyses.  Time
blocks contaminated by RFI are identified using data that are not
corrected for dispersive delay (i.e. DM$ = 0$\,pc\,cm$^{-3}$), in
order to avoid removing astrophysical signals.  The data were
corrected for dispersion using 7292 trial DMs ranging from 0 to
9866.4\,pc\,cm$^{-3}$, generating a time series at each trial.  We
performed Fourier analyses of all the time series to look for periodic
signals using PRESTO's {\tt accelsearch} and detected no significant
signal of a plausible astrophysical origin.

We searched for single pulses in each dispersion-corrected time series
by convolving a template bank of boxcar functions with widths ranging
from 0.13 to 100\,ms.  This optimises the detection of pulses with
durations longer than the native sample time of the data.
Single-pulse events at each DM were identified by applying a
signal-to-noise ratio (S/N) threshold of 5.

These single-pulse events were grouped and ranked using the {\tt
  RRATtrap} sifting algorithm\cite{kkl+15}.  An astrophysical pulse is
detected with maximum S/N at the signal's true DM and is detected with
decreasing S/N at nearby trial DMs.  This is not generally the case
for RFI, whose S/N does not typically peak at a non-zero trial DM.
The {\tt RRATtrap} algorithm ranks candidates based on this DM
behaviour and candidate plots are produced for highly ranked
single-pulse groups.  These plots display the significance of the
pulse as a function of DM and time as well as an image of the signal
as a function of time and observing frequency (e.g. Figure
\ref{fig:bursts}).  The resulting plots were inspected for
astrophysical signals, and pulses were found at a DM of $\sim
559$\,pc\,cm$^{-3}$ at a sky position consistent with the discovery
position of \frb \cite{sch+14}. It is possible that the analysed data
contain weaker bursts, which cannot be reliably identified because
their S/N is too low to distinguish them from RFI or statistical
noise.  If, in the future, the bursts are shown to have an underlying
periodicity, then this would enable a deeper search for weak bursts.

\subsubsection*{Timing analysis of burst arrival times}

Using several approaches, we searched for an underlying periodicity
matching the arrival times of the eight bursts detected in the 2015
June 2 observing session. There are no significant periodicities
detected through a standard Fast Fourier Transform of the time
series. We then carried out a similar analysis to that routinely used
to detect periodicities in sporadically emitting radio
pulsars\cite{mll+06}. In this analysis, we calculate differences
between all of the burst arrival times and search for the greatest
common denominator of these differences. We found several periods, not
harmonically related, that fit different subsets of bursts within a
tolerance of 1\% of the trial period, but none that fit all of the
bursts. We subsequently calculated residuals for the times-of-arrival
for the eight bursts detected on 2015 June 2 for a range of trial
periods using the pulsar timing packages TEMPO and PINT. We found that
some of the periods returned by the differencing algorithm also
resulted in residuals with root-mean-square (RMS) of less than 1\% of
the trial period. However, there were many non-harmonically related
candidate periods resulting in residuals of a comparable
RMS. Furthermore, given the number of trials necessary for this
search, none of these trial periods was statistically significant. In
addition, due to the small number of detected bursts, and the widths
of the pulses, we were not sensitive to periodicities much shorter
than $\sim$~100~ms as our tolerance for a period match (or acceptable
RMS) becomes a large fraction of the period and there are many
possible fits. The 16-day gap between the 2015 May and June detections
precluded us from including the May bursts in any search for
periodicity in the single pulses.

\subsubsection*{Spectral fitting}

In order to produce the spectra shown in the right panels of
Figure~\ref{fig:bursts}, we corrected each spectrum for the bandpass
of the receiver. We estimated the bandpass by taking the average of
the raw data samples for each frequency channel. We then
median-filtered that average bandpass with a width of 20 channels to
remove the effects of narrow-band RFI and divided the observed
spectrum of each burst by this median-filtered bandpass.  The
band-corrected burst spectra shown in the right sub-panels of
Figure~\ref{fig:bursts} are still somewhat contaminated by RFI,
however.  The bottom and top 10 channels (3.4\,MHz) of the band were
ignored due to roll-off in the receiver response.

To characterise the bandpass-corrected spectrum of each burst, we
applied a power-law model using least-squares fitting.  The power-law
model is described by $S_\nu \propto \nu^{\alpha}$, where $S_\nu$ is
the flux density in a frequency channel, $\nu$ is the observing
frequency, and $\alpha$ is the spectral index.  These measured
spectral indices and their uncertainties are shown in
Table~\ref{tab:bursts}.  We do not include a spectral index value for
Burst 6 because of the RFI in the lower half of the band.  For Bursts
7 and 9, we exclude data below 1250\,MHz, because of RFI contamination.
For Bursts 8 and 10, the power-law model was not a good descriptor,
and therefore no value is reported in Table~\ref{tab:bursts}.

Note, we verified this technique by applying the bandpass correction
to PALFA data of pulsar B1900+01.  The measured spectral index was
calculated for ten, bright single pulses, and the values are
consistent with the published value.

\subsubsection*{Measurement of dispersion measure and index}

We measured the DM for 10 of the 11 bursts and additionally the
dispersion index (dispersive delay $\Delta t \propto \nu^{-\xi}$) for
the brightest two.  The DM and DM index, $\xi$, were calculated with a
least-squares routine using the SIMPLEX and MIGRAD functions from the
CERN MINUIT package (http://www.cern.ch/minuit).  The user specifies
the assumed form of the intrinsic pulse shape, which is then convolved
with the appropriate DM smearing factor. For these fits a boxcar pulse
template was used.

Subbanded pulse profiles for each burst were generated by averaging
blocks of frequency channels (i.e. frequency scrunching).  The number
of subbands generated depended on the S/N of the burst to ensure that
there was sufficient S/N in each subband for the fit to converge.
Subbands with no signal were excluded from the fit. Furthermore, the
data were binned in time to further increase the S/N and reduce the
effects of frequency-dependent flux evolution.  As the true intrinsic
pulse width is not known, each burst was fit with a range of boxcar
widths.  The parameters corresponding to the input template yielding
the cleanest residuals are reported.

The DM value was fit keeping the DM index fixed at 2.0. Note, Burst 7
was too weak and corrupted by RFI to obtain reasonable fits.
Additionally, for the brightest two bursts (8 and 11), we also did a
joint fit of DM and DM index.  The resulting DM index fits were
2.00$\pm$0.02 and 1.999$\pm$0.002 for Burst 8 and 11, respectively.
These values are as expected for radio waves traveling through a cold,
ionised medium

Frequency-dependent pulse profile evolution introduces systematic
biases into the times of arrival in each subband.  These biases
in turn bias the DM determination.  These systematics cannot be
mitigated without an accurate model for the underlying burst shape
versus frequency, which is not available in this case, and is further
complicated by the fact that the burst morphology also changes
randomly from burst to burst.  We estimated the systematic uncertainty
by considering what DM value would produce a delay across our
observing band that is comparable to half the burst width in each
case.

Table~\ref{tab:bursts} presents the results of the fits with the
statistical and systematic uncertainties both quoted.  The DM
estimates do not include barycentric corrections (of order 0.01 to
0.1\,pc\,cm$^{-3}$).  Although \frb\ is close to the ecliptic, the
angular separation from the Sun was always much larger than 10
degrees, and any annual contribution to the DM from the solar wind was
small ($< 10^{-3}$ \,pc\,cm$^{-3}$)\cite{ychm12,aab+15}.
These effects are, therefore, much smaller than the aforementioned
systematics in modelling the DMs of the bursts.

The $\pm 1$-$\sigma$ range of DMs for the 10 new bursts is
$558.1\pm3.3$\,pc\,cm$^{-3}$, consistent with the discovery
value\cite{sch+14}, $557.4\pm2.0$\,pc\,cm$^{-3}$.  The DMs and DM
index reported here and previously\cite{sch+14} were calculated using
different methods. These two approaches fit for different free
parameters, so different co-variances between parameters may result in
slightly different values. Also, different time and frequency
resolutions were used. Nonetheless, the Burst 1 parameters quoted here
and previously\cite{sch+14} are consistent within the uncertainties.
The consistency of the DMs is conclusive evidence that a single source
is responsible for the events.

\subsubsection*{The role of interstellar scattering}

Some FRBs have shown clear evidence for multi-path propagation from
scattering by the intervening interstellar or extragalactic material
along the line of sight\cite{tsb+13,rsj15,cpk+15,mls+15a}.  However,
the burst profiles from \frb\ show no obvious evidence for asymmetry
from multi-path propagation.  An upper bound on the pulse broadening
time from Burst 1 is 1.5\,ms at 1.5\,GHz\cite{sch+14}.  Using the
NE2001 model for a source far outside the Galaxy, the expected pulse
broadening is $\sim 20 \rm \mu s \, \nu^{-4.4}$ with $\nu$ in GHz, an
order of magnitude smaller than the $\sim 2$~ms pulse widths and
$\sim0.7$~ms intra-channel dispersion smearing.  The features of the
spectra can not be explained by diffractive interstellar
scintillations; the predicted scintillation bandwidth for \frb\ is
$\sim$50 kHz at 1.5 GHz, which is unresolved by the 0.34 MHz frequency
channels of our data.  We would, therefore, also not expect to observe
diffractive interstellar scintillation in our bursts.  Additional
scattering occurring in a host galaxy and the intergalactic medium is
at a level below our ability to detect.  However, observations at
frequencies below 1.5\,GHz may reveal pulse
broadening that is not substantially smaller than the upper bound if
we use as a guide the observed pulse broadening from other
FRBs\cite{tsb+13,rsj15,cpk+15,mls+15a}.  Future observations that
quantify diffractive interstellar scintillations can provide
constraints on the location of extragalactic scattering plasma
relative to the source, as demonstrated for FRB~110523\cite{mls+15a}.

The upper bound on pulse broadening for \frb\ implies that the
apparent, scattered source size for radio waves incident on the Milky
Way's ISM is small enough so that refractive interstellar
scintillation (RISS) from the ISM is expected.  For the line of sight
to \frb\ we use the NE2001 model to estimate an effective
scattering-screen distance of $\sim 2$\,kpc from Earth and a
scattering diameter of 6\,mas.  The implied length scale for
phase-front curvature is then $l_{\rm RISS} \sim 2$\,kpc$\times$6\,mas
$= 12$\,AU.  For an effective, nominal velocity, $V_{\rm eff} =
100V_{100}$\,km~s$^{-1}$, the expected RISS time scale is $\Delta
t_{\rm RISS} = l_{\rm ISS} / V_{\rm eff} = 215 \, V_{100}^{-1} \,
\nu_{\rm 1\,GHz}^{-2.2}\, ~{\rm days}$.  At 1.5\,GHz and an effective
velocity due to Galactic rotation, which is about 200\,km~s$^{-1}$ in
the direction of \frb, RISS time scales of 20--40 days are expected.
Modulation from RISS can be several tens of percent\cite{r90}.  This
could play a role in the detections of bursts in 2015 mid-May and 2015
June and their absence in 2015 early-May and at other epochs.
However, the solar system and the ionised medium have the same
Galactic rotation, so the effective velocity could be smaller than 100
km~s$^{-1}$, leading to longer RISS time scales.
 
\subsubsection*{Data availability}

The beam positions used in Figure 1 are available as a text file. The
data of the bursts used to generate Figure 2 are provided as a text
file.

\subsubsection*{Code availability}

The code used to analyse the data is available at the following sites:
\\PRESTO (https://github.com/scottransom/presto), RRATtrap
(https://github.com/ckarako/RRATtrap), TEMPO
(http://tempo.sourceforge.net/), and PINT
(http://github.com/nanograv/PINT). \\ \\


\clearpage


\section*{Extended Data}

\setcounter{table}{0}


\captionsetup[table]{name=Extended Data Table}

\begin{table}
\caption{\footnotesize
{\bf Arecibo \frb\ discovery and follow-up observations.}  The observing setup of these observations is described in the Methods.}
\begin{center}
\scriptsize
\begin{tabular}{llllllll}
\hline
\hline
UTC Date         & Project       & Position & Receiver & Frequency & Backend & Dwell time & \# Bursts \\
			&		   &		     &  & (GHz) 	  & 		    & (s) & \\
\hline
\multicolumn{8}{c}{Survey discovery observations presented in Spitler et al. (2014)} \\
\hline
{\bf 2012-11-02} & p2030 & {\bf FRBDISC} & ALFA    & 1.4     & Mocks        & 200\        & {\bf 1 (Beam4)}   \\
\hline
2012-11-04      & p2030 & FRBDISC & ALFA    & 1.4     & Mocks        & 200\        & 0   \\
\hline
\multicolumn{8}{c}{Follow-up observations presented in Spitler et al. (2014)} \\
\hline
2013-12-09      & p2886 & FRBGRID1a & ALFA       & 1.4     & Mocks        & 2700      & 0     \\
2013-12-09      & p2886 & FRBGRID2a & ALFA       & 1.4     & Mocks        & 970        & 0     \\
2013-12-09      & p2886 & FRBGRID2a & ALFA       & 1.4     & Mocks        & 1830      & 0     \\
\hline
2013-12-10      & p2886 & FRBGRID3a & ALFA       & 1.4     & Mocks        & 2700      & 0     \\
2013-12-10      & p2886 & FRBDISC   & 327-MHz & 0.327 & PUPPI       & 2385      & 0     \\
\hline
\multicolumn{8}{c}{Follow-up observations presented here for the first time} \\
\hline
2015-05-02      & p2886 & FRBDISC   & L-wide     & 1.4      & PUPPI       & 7200 & 0     \\
\hline
2015-05-03      & p2886 & FRBGRID1b & ALFA      & 1.4      & Mocks        & 1502      & 0   \\
2015-05-03      & p2886 & FRBGRID2b & ALFA      & 1.4      & Mocks        & 1502      & 0   \\
2015-05-03      & p2886 & FRBGRID3b & ALFA      & 1.4      & Mocks        & 343        & 0   \\
2015-05-03      & p2886 & FRBGRID3b & ALFA      & 1.4      & Mocks        & 1502      & 0   \\
2015-05-03      & p2886 & FRBGRID1b & ALFA      & 1.4      & Mocks        & 921        & 0   \\
\hline
2015-05-05      & p2886 & FRBGRID1b & ALFA      & 1.4      & Mocks        & 1002      & 0   \\
2015-05-05      & p2886 & FRBGRID2b & ALFA      & 1.4      & Mocks        & 1002      & 0   \\
2015-05-05      & p2886 & FRBGRID3b & ALFA      & 1.4      & Mocks        & 1002      & 0   \\
2015-05-05      & p2886 & FRBGRID4b & ALFA      & 1.4      & Mocks        & 1002      & 0   \\
2015-05-05      & p2886 & FRBGRID5b & ALFA      & 1.4      & Mocks        & 1002      & 0   \\
2015-05-05      & p2886 & FRBGRID6b & ALFA      & 1.4      & Mocks        & 1002      & 0   \\
\hline
2015-05-09      & p2886 & FRBGRID1b & ALFA      & 1.4      & Mocks        & 1002      & 0   \\
2015-05-09      & p2886 & FRBGRID2b & ALFA      & 1.4      & Mocks        & 1002      & 0   \\
2015-05-09      & p2886 & FRBGRID3b & ALFA      & 1.4      & Mocks        & 1002      & 0   \\
2015-05-09      & p2886 & FRBGRID4b & ALFA      & 1.4      & Mocks        & 1002      & 0   \\
2015-05-09      & p2886 & FRBGRID5b & ALFA      & 1.4      & Mocks        & 1002      & 0   \\
2015-05-09      & p2886 & FRBGRID6b & ALFA      & 1.4      & Mocks        & 1002      & 0   \\
2015-05-09      & p2886 & FRBGRID6b & ALFA      & 1.4      & Mocks        & 425        & 0   \\
\hline
2015-05-17      & p2886 & FRBGRID1b & ALFA      & 1.4      & Mocks        & 1002      & 0   \\
{\bf 2015-05-17} & p2886 & {\bf FRBGRID2b} & ALFA      & 1.4      & Mocks        & 1002      & {\bf 2 (Beam6)}   \\
2015-05-17      & p2886 & FRBGRID3b & ALFA      & 1.4      & Mocks        & 1002      & 0   \\
2015-05-17      & p2886 & FRBGRID4b & ALFA      & 1.4      & Mocks        & 707        & 0   \\
2015-05-17      & p2886 & FRBGRID4b & ALFA      & 1.4      & Mocks        & 391        & 0   \\
2015-05-17      & p2886 & FRBGRID5b & ALFA      & 1.4      & Mocks        & 1002      & 0   \\
2015-05-17      & p2886 & FRBGRID6b & ALFA      & 1.4      & Mocks        & 1002      & 0   \\
\hline
2015-06-02      & p2886 & FRBGRID1b & ALFA      & 1.4      & Mocks        & 1002      & 0   \\
{\bf 2015-06-02} & p2886 & {\bf FRBGRID2b} & ALFA      & 1.4      & Mocks        & 1002      & {\bf 2 (Beam6)}   \\
2015-06-02      & p2886 & FRBGRID3b & ALFA      & 1.4      & Mocks        & 1002      & 0   \\
2015-06-02      & p2886 & FRBGRID4b & ALFA      & 1.4      & Mocks        & 1002      & 0   \\
2015-06-02      & p2886 & FRBGRID5b & ALFA      & 1.4      & Mocks        & 1002      & 0   \\
{\bf 2015-06-02} & p2886 & {\bf FRBGRID6b} & ALFA      & 1.4      & Mocks        & 1002      & {\bf 6 (Beam0)} \\
2015-06-02      & p2886 & FRBGRID6b & ALFA      & 1.4      & Mocks        & 300        & 0  \\
\hline
\hline
\end{tabular}
\end{center}
\label{tab:observations}
\end{table}

\clearpage

\begin{table}
\caption{\footnotesize
{\bf \frb\ gridding positions.}  The J2000 ALFA Beam 0 positions are listed.}
\begin{center}
\scriptsize
\begin{tabular}{lll}
\hline
\hline
Grid ID         & Right Ascension       & Declination \\
\hline
FRBDISC      & $05^h32^m09^s$ & $+33^d05^m13^s$ \\
FRBGRID1a & $05^h32^m16^s$ & $+33^d05^m39^s$ \\
FRBGRID2a & $05^h32^m22^s$ & $+33^d03^m06^s$ \\
FRBGRID3a & $05^h32^m11^s$ & $+33^d03^m06^s$ \\
FRBGRID1b & $05^h32^m10^s$ & $+33^d05^m13^s$ \\
FRBGRID2b & $05^h32^m24^s$ & $+33^d05^m13^s$ \\
FRBGRID3b & $05^h31^m55^s$ & $+33^d05^m13^s$ \\
FRBGRID4b & $05^h32^m10^s$ & $+33^d08^m13^s$ \\
FRBGRID5b & $05^h32^m24^s$ & $+33^d08^m13^s$ \\
FRBGRID6b & $05^h31^m55^s$ & $+33^d08^m13^s$ \\
\hline
\hline
\end{tabular}
\end{center}
\label{tab:gridpoints}
\end{table}

\end{document}